\documentclass[12pt]{article}

\usepackage{feynarts}
\usepackage{epsfig}

\makeatletter

\def\paragraph{\@startsection{paragraph}{4}{\z@}{+2.00ex plus
 +1ex minus +.2ex}{1.5ex plus .2ex}{\it\normalsize}}

\def\section{\@startsection {section}{1}{\z@}{+3.0ex plus +1ex minus
  +.2ex}{2.3ex plus .2ex}{\normalsize\bf\boldmath}}
\def\subsection{\@startsection{subsection}{2}{\z@}{+2.5ex plus +1ex
minus +.2ex}{1.5ex plus .2ex}{\normalsize\bf\boldmath}}
\def\subsubsection{\@startsection{subsubsection}{3}{\z@}{+3.25ex plus
 +1ex minus +.2ex}{1.5ex plus .2ex}{\normalsize\it}}

\expandafter\ifx\csname mathrm\endcsname\relax\def\mathrm#1{{\rm #1}}\fi

\newcounter{saveeqn}

\@addtoreset{equation}{section}

\newcount\@tempcntc
\def\@citex[#1]#2{\if@filesw\immediate\write\@auxout{\string\citation{#2}}\fi
  \@tempcnta\z@\@tempcntb\m@ne\def\@citea{}\@cite{\@for\@citeb:=#2\do
    {\@ifundefined
       {b@\@citeb}{\@citeo\@tempcntb\m@ne\@citea
        \def\@citea{,\penalty\@m\ }{\bf ?}\@warning
       {Citation `\@citeb' on page \thepage \space undefined}}%
    {\setbox\z@\hbox{\global\@tempcntc0\csname
b@\@citeb\endcsname\relax}%
     \ifnum\@tempcntc=\z@ \@citeo\@tempcntb\m@ne
       \@citea\def\@citea{,\penalty\@m}
       \hbox{\csname b@\@citeb\endcsname}%
     \else
      \advance\@tempcntb\@ne
      \ifnum\@tempcntb=\@tempcntc
      \else\advance\@tempcntb\m@ne\@citeo
      \@tempcnta\@tempcntc\@tempcntb\@tempcntc\fi\fi}}\@citeo}{#1}}

\def\@citeo{\ifnum\@tempcnta>\@tempcntb\else\@citea
  \def\@citea{,\penalty\@m}%
  \ifnum\@tempcnta=\@tempcntb\the\@tempcnta\else
   {\advance\@tempcnta\@ne\ifnum\@tempcnta=\@tempcntb \else
\def\@citea{--}\fi
    \advance\@tempcnta\m@ne\the\@tempcnta\@citea\the\@tempcntb}\fi\fi}

\def\asymp#1%
{\mathrel{\raisebox{-.4em}{$\widetilde{\scriptstyle #1}$}}}

\def\Nequal#1%
{\mathrel{\raisebox{-.5em}{$\stackrel{=}{\scriptstyle\rm#1}$}}}
\newcommand{\dsl}[1]{\not \hspace{-0.7mm}#1}
\def\dsl{\mathpalette\make@slash}
\def\make@slash#1#2{\setbox\z@\hbox{$#1#2$}%
  \hbox to 0pt{\hss$#1/$\hss\kern-\wd0}\box0}

\def\beq{\begin{equation}}
\def\eeq{\end{equation}}
\def\beqar{\begin{eqnarray}}
\def\eeqar{\end{eqnarray}}
\def\barr#1{\begin{array}{#1}}
\def\earr{\end{array}}
\def\bfi{\begin{figure}}
\def\efi{\end{figure}}
\def\btab{\begin{table}}
\def\etab{\end{table}}
\def\bce{\begin{center}}
\def\ece{\end{center}}
\def\nn{\nonumber}
\def\disp{\displaystyle}
\def\text{\textstyle}

\def\ga{\gamma}
\def\de{\delta}
\def\De{\Delta}

\def\si{\sigma}

\def\refeq#1{\mbox{(\ref{#1})}}

\def\reffi#1{\mbox{Figure~\ref{#1}}}

\def\refta#1{\mbox{Table~\ref{#1}}}

\def\refse#1{\mbox{Section~\ref{#1}}}

\def\citere#1{\mbox{Ref.~\cite{#1}}}
\def\citeres#1{\mbox{Refs.~\cite{#1}}}

\newcommand{\GeV}{\unskip\,\mathrm{GeV}}
\newcommand{\MeV}{\unskip\,\mathrm{MeV}}

\newcommand{\rd}{{\mathrm{d}}}

\newcommand{\M}{{\cal{M}}}

\def\mathswitchr#1{\relax\ifmmode{\mathrm{#1}}\else$\mathrm{#1}$\fi}

\newcommand{\PW}{\mathswitchr W}
\newcommand{\Pw}{\mathswitchr w}
\newcommand{\PZ}{\mathswitchr Z}

\newcommand{\PH}{\mathswitchr H}
\newcommand{\Pe}{\mathswitchr e}
\newcommand{\Pp}{\mathswitchr p}
\newcommand{\Pn}{\mathswitchr n}

\newcommand{\Pd}{\mathswitchr d}

\newcommand{\Pu}{\mathswitchr u}

\newcommand{\Ps}{\mathswitchr s}

\newcommand{\Pc}{\mathswitchr c}

\newcommand{\Pb}{\mathswitchr b}

\newcommand{\Pt}{\mathswitchr t}

\def\mathswitch#1{\relax\ifmmode#1\else$#1$\fi}

\newcommand{\MW}{\mathswitch {M_\PW}}

\newcommand{\MZ}{\mathswitch {M_\PZ}}
\newcommand{\MH}{\mathswitch {M_\PH}}
\newcommand{\Me}{\mathswitch {m_\Pe}}

\newcommand{\Md}{\mathswitch {m_\Pd}}
\newcommand{\Mu}{\mathswitch {m_\Pu}}
\newcommand{\Ms}{\mathswitch {m_\Ps}}
\newcommand{\Mc}{\mathswitch {m_\Pc}}
\newcommand{\Mb}{\mathswitch {m_\Pb}}
\newcommand{\Mt}{\mathswitch {m_\Pt}}

\newcommand{\sw}{\mathswitch {s_\Pw}}
\newcommand{\cw}{\mathswitch {c_\Pw}}

\newcommand{\GF}{\mathswitch {G_\mu}}

\def\solid{\raise.9mm\hbox{\protect\rule{1.1cm}{.2mm}}}
\def\dash{\raise.9mm\hbox{\protect\rule{2mm}{.2mm}}\hspace*{1mm}}

\newcommand{\ct}{\mathrm{ct}}

\newcommand{\NC}{{\mathrm{NC}}}
\newcommand{\CC}{{\mathrm{CC}}}
\newcommand{\iso}{{\mathrm{iso}}}

\newcommand{\brem}{{\mathrm{brem}}}

\newcommand{\onel}{{\mathrm{one-loop}}}

\newcommand{\cut}{{\mathrm{cut}}}

\newcommand{\had}{{\mathrm{had}}}
\newcommand{\phot}{{\mathrm{phot}}}
\newcommand{\recomb}{{\mathrm{recomb}}}
\newcommand{\LAB}{{\mathrm{LAB}}}

\newcommand{\virt}{{\mathrm{virt}}}
\newcommand{\real}{{\mathrm{real}}}

\newcommand{\coll}{{\mathrm{coll}}}

\renewcommand{\min}{{\mathrm{min}}}
\renewcommand{\max}{{\mathrm{max}}}

\def\Re{\mathop{\mathrm{Re}}\nolimits}

\def\draftdate{\relax}
\def\mda{\relax}
\def\mua{\relax}
\def\mla{\relax}
\def\draft{
\def\thtystars{******************************}
\def\sixtystars{\thtystars\thtystars}
\typeout{}
\typeout{\sixtystars**}
\typeout{* Draft mode!
         For final version remove \protect\draft\space in source file *}
\typeout{\sixtystars**}
\typeout{}
\def\draftdate{\today}
\def\mua{\marginpar[\boldmath\hfil$\uparrow$]%
                   {\boldmath$\uparrow$\hfil}%
                    \typeout{marginpar: $\uparrow$}\ignorespaces}
\def\mda{\marginpar[\boldmath\hfil$\downarrow$]%
                   {\boldmath$\downarrow$\hfil}%
                    \typeout{marginpar: $\downarrow$}\ignorespaces}
\def\mla{\marginpar[\boldmath\hfil$\rightarrow$]%
                   {\boldmath$\leftarrow $\hfil}%
                    \typeout{marginpar: $\leftrightarrow$}\ignorespaces}
\def\Mua{\marginpar[\boldmath\hfil$\Uparrow$]%
                   {\boldmath$\Uparrow$\hfil}%
                    \typeout{marginpar: $\uparrow$}\ignorespaces}
\def\Mda{\marginpar[\boldmath\hfil$\Downarrow$]%
                   {\boldmath$\Downarrow$\hfil}%
                    \typeout{marginpar: $\downarrow$}\ignorespaces}
\def\Mla{\marginpar[\boldmath\hfil$\Rightarrow$]%
                   {\boldmath$\Leftarrow $\hfil}%
                    \typeout{marginpar: $\leftrightarrow$}\ignorespaces}
\overfullrule 5pt
\oddsidemargin -15mm
\marginparwidth 29mm
}

\def\stars{\strut\leaders\hbox{*}\hfill\strut}
\def\starline{\hfil\strut\hfil\hbox to \textwidth {\stars}\hfil}

\begin{document}
\thispagestyle{empty}
\def\thefootnote{\fnsymbol{footnote}}
\setcounter{footnote}{1}
\null
\draftdate\hfill MPP-2003-48 \\
\strut\hfill PSI-PR-03-16\\
\strut\hfill hep-ph/0310364
\vfill
\begin{center}
{\Large \bf\boldmath{
Electroweak radiative corrections to deep-inelastic neutrino scattering
--- implications for NuTeV ?}
\par} \vskip 2.5em
\vspace{1cm}

{\large
{\sc K.-P.~O.~Diener$^1$, S.~Dittmaier$^2$ and W.~Hollik$^2$} } \\[1cm]
$^1$ {\it Paul-Scherrer-Institut, W\"urenlingen und Villigen\\
CH-5232 Villigen PSI, Switzerland} \\[0.5cm]
$^2$ {\it Max-Planck-Institut f\"ur Physik 
(Werner-Heisenberg-Institut) \\
D-80805 M\"unchen, Germany}
\par \vskip 1em
\end{center}\par
\vskip 2cm {\bf Abstract:} \par 
We calculate the ${\cal O}(\alpha)$ electroweak corrections to charged-
and neutral-current deep-inelastic neutrino scattering off an isoscalar target.
The full one-loop-corrected cross sections, 
including hard photonic corrections, 
are evaluated and compared to an earlier result which was used in
the NuTeV analysis.
In particular, we compare results that differ in input-parameter 
scheme, treatment of real photon radiation and factorization scheme.
The associated shifts in the theoretical prediction for the ratio of 
neutral- and
charged-current cross sections can be larger than the experimental accuracy of 
the NuTeV result. 
\par
\vskip 3cm
\noindent
October 2003
\null
\setcounter{page}{0}
\clearpage
\def\thefootnote{\arabic{footnote}}
\setcounter{footnote}{0}

\clearpage

\section{Introduction}
\label{sec:intro}

Deep-inelastic neutrino scattering has been analyzed in the
NuTeV experiment \cite{Zeller:2001hh} with a rather high precision.
In detail, the neutral- to charged-current cross-section 
ratios~\cite{LlewellynSmith:ie}
\beq
R^\nu = \frac{\si_\NC^\nu(\nu_\mu N\to\nu_\mu X)}
{\si_\CC^\nu(\nu_\mu N\to\mu^- X)}, 
\qquad
R^{\bar\nu} = \frac{\si_\NC^{\bar\nu}(\bar\nu_\mu N\to\bar\nu_\mu X)}%
{\si_\CC^{\bar\nu}(\bar\nu_\mu N\to\mu^+ X)}
\label{eq:Rnu}
\eeq
have been measured to an accuracy of about 0.2\% and 0.4\%, respectively.
In addition, the quantity
\beq
R^- = \frac{\si_\NC^\nu(\nu_\mu N\to\nu_\mu X)
-\si_\NC^{\bar\nu}(\bar\nu_\mu N\to\bar\nu_\mu X)}
{\si_\CC^\nu(\nu_\mu N\to\mu^- X)-\si_\CC^{\bar\nu}(\bar\nu_\mu N\to\mu^+ X)},
\eeq
as proposed by Paschos and Wolfenstein \cite{Paschos:1972kj},
has been considered.
As a central result, the NuTeV collaboration has translated their
measurement into a value for the on-shell 
weak mixing angle, $\sin^2\theta_\PW=1-\MW^2/\MZ^2$, which can, 
thus, be viewed as an independent (but
rather indirect) determination of the W- to Z-boson mass ratio.
The NuTeV result on $\sin^2\theta_\PW$ is, however, about $3\si$ away from the
result obtained from the global fit \cite{Grunewald:2003ij}
of the Standard Model (SM) to the electroweak precision data.

The $3\si$ difference of the NuTeV result to the SM global fit to data,
together with the fact that the evaluation of the NuTeV measurement is
very involved, triggered some discussion about the reliability of the
NuTeV result. A detailed discussion of various experimental as well as
theoretical issues can be found in \citere{McFarland:2003jw}, 
where in particular QCD corrections (see also
\citere{Davidson:2001ji} and references therein), 
charm-mass effects, and 
uncertainties from the parton 
distribution functions have been considered. 
According to \citere{McFarland:2003jw}, none of these 
issues seems to be a candidate for an error or an underestimate 
of uncertainties in the analysis. However, the authors of 
\citere{McFarland:2003jw} pointed out that the inclusion of electroweak radiative corrections,
which influences the result significantly, is based on a single
calculation \cite{Bardin:1986bc} only%
\footnote{Electroweak radiative corrections to neutral-current-induced 
neutrino deep-inelastic scattering were also investigated in
\citere{Marciano:pb}, but the corresponding hadronic cross sections
were not evaluated numerically.}
and that a careful recalculation of these corrections is desirable. 

In this paper we present such a calculation of the ${\cal O}(\alpha)$ 
electroweak corrections to neutral- and charged-current deep-inelastic 
neutrino scattering off an isoscalar target. 
Apart from a different set of parton densities and input parameters 
the most important difference between our calculation and the result of 
\citere{Bardin:1986bc} lies in the treatment of initial-state mass
singularities due to collinear radiation of a photon from the incoming quark. 

The input parameters used for the numerical analysis of \citere{Bardin:1986bc}
are not completely specified and, 
in order to fix all free parameters of the electroweak SM, 
need to be supplemented by fermion masses that we obtained from references
quoted within \citere{Bardin:1986bc}.
A comparison of numerical results
should therefore be interpreted with care.
In any case, the effect on the electroweak corrections on the NuTeV results
should be further analyzed.
We will provide a Fortran code that could be used for an update of the
NuTeV data analysis.

The paper is organized as follows.
In \refse{sec:notation} we set our conventions, define the necessary
kinematical quantities, and consider the lowest-order cross sections.
Section~\ref{se:RCs} contains an overview over the structure of the
calculated electroweak corrections, a description of the calculational 
framework, and explicit results on some specific one-loop corrections;
particular attention is paid to the treatment of collinear photon
radiation and the related fermion-mass singularities.
Our numerical results are presented in \refse{se:results}. 
Section~\ref{se:summary} contains a short summary. 

\section{Conventions and lowest-order cross sections}
\label{sec:notation}

\begin{figure}
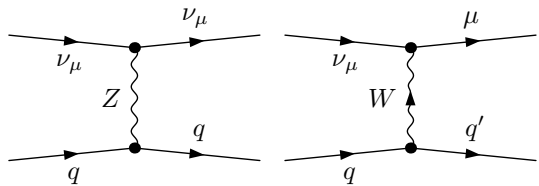

\centerline{\footnotesize  
\unitlength=1bp%
\begin{feynartspicture}(216,104)(2,1)
\FADiagram{}
\FAProp(0.,15.)(10.,14.)(0.,){/Straight}{1}
\FALabel(4.84577,13.4377)[t]{$\nu_\mu$}
\FAProp(0.,5.)(10.,6.)(0.,){/Straight}{1}
\FALabel(5.15423,4.43769)[t]{$q$}
\FAProp(20.,15.)(10.,14.)(0.,){/Straight}{-1}
\FALabel(14.8458,15.5623)[b]{$\nu_\mu$}
\FAProp(20.,5.)(10.,6.)(0.,){/Straight}{-1}
\FALabel(15.1542,6.56231)[b]{$q$}
\FAProp(10.,14.)(10.,6.)(0.,){/Sine}{0}
\FALabel(8.93,10.)[r]{$Z$}
\FAVert(10.,14.){0}
\FAVert(10.,6.){0}
\FADiagram{}
\FAProp(0.,15.)(10.,14.)(0.,){/Straight}{1}
\FALabel(4.84577,13.4377)[t]{$\nu_\mu$}
\FAProp(0.,5.)(10.,6.)(0.,){/Straight}{1}
\FALabel(5.15423,4.43769)[t]{$q$}
\FAProp(20.,15.)(10.,14.)(0.,){/Straight}{-1}
\FALabel(14.8458,15.5623)[b]{$\mu$}
\FAProp(20.,5.)(10.,6.)(0.,){/Straight}{-1}
\FALabel(15.1542,6.56231)[b]{$q'$}
\FAProp(10.,14.)(10.,6.)(0.,){/Sine}{-1}
\FALabel(8.93,10.)[r]{$W$}
\FAVert(10.,14.){0}
\FAVert(10.,6.){0}
\end{feynartspicture}
} \vspace*{-2em}
\caption{Lowest-order diagrams for 
$\nu_\mu q \to \nu_\mu q$ and 
$\nu_\mu q \to \mu^- q'$ scattering}
\label{fig:LOdiags}
\end{figure}

We consider the neutral-current (NC) and charged-current (CC) parton processes
\beqar
\mbox{NC:} \quad \nu_\mu(p_l) + q(p_q) &\;\to\;& \nu_\mu(k_l) + q(k_q), 
\qquad q=\Pu,\Pd,\Ps,\Pc,\bar\Pu,\bar\Pd,\bar\Ps,\bar\Pc,
\label{eq:ncprocesses}
\\
\mbox{CC:} \quad \nu_\mu(p_l) + q(p_q) &\;\to\;& \mu^-(k_l) + q'(k_q),
\qquad q=\Pd,\Ps,\bar\Pu,\bar\Pc,
\quad q'=\Pu,\Pc,\bar\Pd,\bar\Ps,
\label{eq:ccprocesses}
\eeqar
and the processes with all particles replaced by their antiparticles.
The assignment of momenta is indicated in parentheses.
At lowest order the former processes proceed via Z-boson exchange, the latter
via W-boson exchange; the corresponding Feynman diagrams
are shown in \reffi{fig:LOdiags}.
The diagrams and results for incoming antiquarks can be obtained 
from those of incoming quarks by crossing the external quark lines.
CP symmetry implies that
the (parton) cross sections for $\nu_\mu$--quark and $\nu_\mu$--antiquark
scattering are equal to the ones of 
$\bar\nu_\mu$--antiquark and $\bar\nu_\mu$--quark scattering,
respectively.
In the following we neglect the muon mass and the parton masses
whenever possible. The quark masses are only kept as regulators
for mass singularities, which appear as mass logarithms in the
(photonic) corrections; the issue of quark-mass logarithms will be
discussed below in detail.
The muon mass is not only required as regulator for collinear
final-state radiation but also for a proper description of 
forward scattering in the (loop-induced) $\gamma\nu_\mu\bar\nu_\mu$ vertex,
as described below as well.
The Mandelstam variables of the partonic processes are defined as
\beqar
s&=&(p_l+p_q)^2 = 4E^2, \nn\\
t&=&(p_l-k_l)^2 = -2E^2(1-\cos\theta), \nn\\
u&=&(p_q-k_l)^2 = -2E^2(1+\cos\theta),
\eeqar
where $E$ and $\theta$ denote the beam energy and scattering angle
in the partonic centre-of-mass (CM) frame.
Moreover, we introduce the variable
\beq
y = -\frac{t}{s} = \frac{1}{2}(1-\cos\theta).
\eeq

The momentum of the incoming (anti-)quark $q$ is related to the
total nucleon momentum $P_N$ by the usual scaling relation
\beq
p_q^\mu = x P_N^\mu,
\eeq
which holds in the CM frame of the partonic system where the nucleon
mass $M_N$ is negligible. The variable $x$ is the usual momentum
fraction, restricted by $0<x<1$.
Since deep-inelastic neutrino scattering usually is accessed 
in fixed-target
experiments (as NuTeV), we take the energy $E_\nu^{\LAB}$ of the
incoming (anti-)neutrino beam to define the incoming momentum $p_l$
in the rest frame of the nucleon, called LAB in the following. 
Thus, the scattering energy $E$ and the variable $s$ are given by
\beq
s = 4E^2 = 2x M_N E_\nu^{\LAB},
\label{eq:s_EnuLAB}
\eeq
up to terms of higher order in the nucleon mass.
Neglecting the nucleon mass, there is also a simple relation
between the scattering angle $\theta$
in the partonic CM frame (or equivalently $y$) and 
the energy $E_{\had}^{\LAB}$ of the outgoing quark in the LAB
frame,
\beq
E_{\had}^{\LAB} = \frac{E_\nu^{\LAB}}{2}(1-\cos\theta) = y E_\nu^{\LAB}.
\eeq
$E_{\had}^{\LAB}$ can be identified as
the energy deposited by
hadrons in the detector.
More generally, a momentum with components $K^\mu$ in the CM frame
receives the following components in the LAB system,
\beqar
K^\mu_{\LAB} &=& 
\Bigl(\gamma(K^0+\beta K^3),K^1,K^2,\gamma(\beta K^0+K^3)\Bigr)
\nn\\
&& \mbox{with} \quad \gamma\sim\sqrt{E_\nu^{\LAB}/(2x M_N)}, \quad \beta\to1.
\label{eq:boost}
\eeqar
Note that the asymptotic expressions for $\gamma$ and $\beta$ are fully
sufficient and can be consistently used; deviations from the full Lorentz
boost are suppressed by nucleon-mass terms.

The lowest-order matrix elements corresponding to the tree diagrams
of \reffi{fig:LOdiags} can be easily worked out; the result is
\beqar
\M_{\NC,0}^\tau &=& \frac{e^2 g^\tau_{qqZ} g^-_{\nu_\mu\nu_\mu Z}}{t-\MZ^2} \;
\Big[\bar u(k_l) \, \gamma^\rho \, \omega_- u(p_l)\Big] \,
\Big[\bar u(k_q) \, \gamma_\rho \, \omega_\tau u(p_q)\Big],
\nn\\
\M_{\CC,0}^\tau &=& \frac{e^2 g^-_{q'qW} g^-_{\nu_\mu\mu W}}{t-\MW^2} \;
\Big[\bar u(k_l) \, \gamma^\rho \, \omega_- u(p_l)\Big] \,
\Big[\bar u(k_{q'}) \, \gamma_\rho \, \omega_- u(p_q)\Big],
\label{eq:m0}
\eeqar 
where $\omega_\pm=\frac{1}{2}(1\pm\gamma_5)$ denote the chirality projectors
and an obvious notation for the Dirac spinors $u(p_q)$, etc., is used.
The index $\tau=\pm$ 
indicates the chirality of the incoming quark.
The coupling factors are given by
\beq
g^\tau_{q'qW} = \disp \frac{V_{q'q}}{\sqrt{2}\sw}\,\de_{\tau-},
\qquad
g^\tau_{\nu_\mu\mu W} = \disp \frac{1}{\sqrt{2}\sw}\,\de_{\tau-},
\qquad
g^\tau_{ffZ} = -\frac{\sw}{\cw}Q_f+\frac{I_f^3}{\cw\sw}\,\de_{\tau-},
\eeq
where $Q_f$ and $I_f^3=\pm1/2$ are the relative charge and the third
component of the weak isospin of fermion $f$, respectively.
The weak mixing angle is fixed by the mass ratio $\MW/\MZ$,
according to the on-shell condition
$\sin^2 \theta_W\equiv s_W^2 =1-c_W^2 =1-\MW^2/\MZ^2$.
Note that the CKM matrix element for the $q'q$ transition,
$V_{q'q}$, appears only as global factor $|V_{q'q}|^2$
in the CC cross section. Thus, in the limit of vanishing final-state
quark masses, the factors $|V_{q'q}|^2$ add up to one after the sum over
the flavours of $q'$ is taken. 

\section{Calculation of radiative corrections}
\label{se:RCs}

\subsection{Overview and calculational framework}
\label{se:calcframe}

\begin{figure}
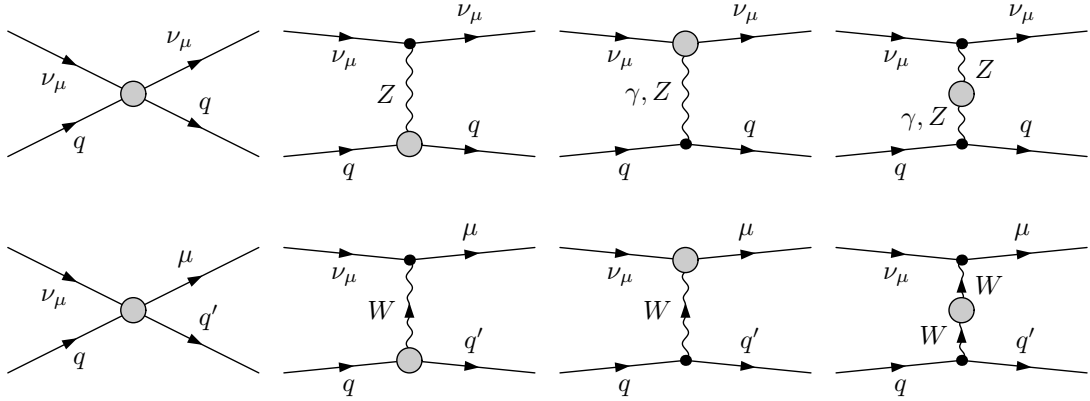

\centerline{\footnotesize  
\unitlength=1bp%
\begin{feynartspicture}(532,104)(4,1)
\FADiagram{}
\FAProp(0.,15.)(10.,10.)(0.,){/Straight}{1}
\FALabel(4.78682,11.5936)[tr]{$\nu_\mu$}
\FAProp(0.,5.)(10.,10.)(0.,){/Straight}{1}
\FALabel(5.21318,6.59364)[tl]{$q$}
\FAProp(20.,15.)(10.,10.)(0.,){/Straight}{-1}
\FALabel(14.7868,13.4064)[br]{$\nu_\mu$}
\FAProp(20.,5.)(10.,10.)(0.,){/Straight}{-1}
\FALabel(15.2132,8.40636)[bl]{$q$}
\FAVert(10.,10.){-1}
\FADiagram{}
\FAProp(0.,15.)(10.,14.)(0.,){/Straight}{1}
\FALabel(4.84577,13.4377)[t]{$\nu_\mu$}
\FAProp(0.,5.)(10.,6.)(0.,){/Straight}{1}
\FALabel(5.15423,4.43769)[t]{$q$}
\FAProp(20.,15.)(10.,14.)(0.,){/Straight}{-1}
\FALabel(14.8458,15.5623)[b]{$\nu_\mu$}
\FAProp(20.,5.)(10.,6.)(0.,){/Straight}{-1}
\FALabel(15.1542,6.56231)[b]{$q$}
\FAProp(10.,14.)(10.,6.)(0.,){/Sine}{0}
\FALabel(8.93,10.)[r]{$Z$}
\FAVert(10.,14.){0}
\FAVert(10.,6.){-1}
\FADiagram{}
\FAProp(0.,15.)(10.,14.)(0.,){/Straight}{1}
\FALabel(4.84577,13.4377)[t]{$\nu_\mu$}
\FAProp(0.,5.)(10.,6.)(0.,){/Straight}{1}
\FALabel(5.15423,4.43769)[t]{$q$}
\FAProp(20.,15.)(10.,14.)(0.,){/Straight}{-1}
\FALabel(14.8458,15.5623)[b]{$\nu_\mu$}
\FAProp(20.,5.)(10.,6.)(0.,){/Straight}{-1}
\FALabel(15.1542,6.56231)[b]{$q$}
\FAProp(10.,14.)(10.,6.)(0.,){/Sine}{0}
\FALabel(8.93,10.)[r]{$\gamma,Z$}
\FAVert(10.,6.){0}
\FAVert(10.,14.){-1}
\FADiagram{}
\FAProp(0.,15.)(10.,14.)(0.,){/Straight}{1}
\FALabel(4.84577,13.4377)[t]{$\nu_\mu$}
\FAProp(0.,5.)(10.,6.)(0.,){/Straight}{1}
\FALabel(5.15423,4.43769)[t]{$q$}
\FAProp(20.,15.)(10.,14.)(0.,){/Straight}{-1}
\FALabel(14.8458,15.5623)[b]{$\nu_\mu$}
\FAProp(20.,5.)(10.,6.)(0.,){/Straight}{-1}
\FALabel(15.1542,6.56231)[b]{$q$}
\FAProp(10.,10.)(10.,14.)(0.,){/Sine}{0}
\FALabel(11.07,12.)[l]{$Z$}
\FAProp(10.,10.)(10.,6.)(0.,){/Sine}{0}
\FALabel(8.93,8.)[r]{$\gamma,Z$}
\FAVert(10.,14.){0}
\FAVert(10.,6.){0}
\FAVert(10.,10.){-1}
\end{feynartspicture}
} 
\vspace*{-2em}
\centerline{\footnotesize  
\unitlength=1bp%
\begin{feynartspicture}(532,104)(4,1)
\FADiagram{}
\FAProp(0.,15.)(10.,10.)(0.,){/Straight}{1}
\FALabel(4.78682,11.5936)[tr]{$\nu_\mu$}
\FAProp(0.,5.)(10.,10.)(0.,){/Straight}{1}
\FALabel(5.21318,6.59364)[tl]{$q$}
\FAProp(20.,15.)(10.,10.)(0.,){/Straight}{-1}
\FALabel(14.7868,13.4064)[br]{$\mu$}
\FAProp(20.,5.)(10.,10.)(0.,){/Straight}{-1}
\FALabel(15.2132,8.40636)[bl]{$q'$}
\FAVert(10.,10.){-1}
\FADiagram{}
\FAProp(0.,15.)(10.,14.)(0.,){/Straight}{1}
\FALabel(4.84577,13.4377)[t]{$\nu_\mu$}
\FAProp(0.,5.)(10.,6.)(0.,){/Straight}{1}
\FALabel(5.15423,4.43769)[t]{$q$}
\FAProp(20.,15.)(10.,14.)(0.,){/Straight}{-1}
\FALabel(14.8458,15.5623)[b]{$\mu$}
\FAProp(20.,5.)(10.,6.)(0.,){/Straight}{-1}
\FALabel(15.1542,6.56231)[b]{$q'$}
\FAProp(10.,14.)(10.,6.)(0.,){/Sine}{-1}
\FALabel(8.93,10.)[r]{$W$}
\FAVert(10.,14.){0}
\FAVert(10.,6.){-1}
\FADiagram{}
\FAProp(0.,15.)(10.,14.)(0.,){/Straight}{1}
\FALabel(4.84577,13.4377)[t]{$\nu_\mu$}
\FAProp(0.,5.)(10.,6.)(0.,){/Straight}{1}
\FALabel(5.15423,4.43769)[t]{$q$}
\FAProp(20.,15.)(10.,14.)(0.,){/Straight}{-1}
\FALabel(14.8458,15.5623)[b]{$\mu$}
\FAProp(20.,5.)(10.,6.)(0.,){/Straight}{-1}
\FALabel(15.1542,6.56231)[b]{$q'$}
\FAProp(10.,14.)(10.,6.)(0.,){/Sine}{-1}
\FALabel(8.93,10.)[r]{$W$}
\FAVert(10.,6.){0}
\FAVert(10.,14.){-1}
\FADiagram{}
\FAProp(0.,15.)(10.,14.)(0.,){/Straight}{1}
\FALabel(4.84577,13.4377)[t]{$\nu_\mu$}
\FAProp(0.,5.)(10.,6.)(0.,){/Straight}{1}
\FALabel(5.15423,4.43769)[t]{$q$}
\FAProp(20.,15.)(10.,14.)(0.,){/Straight}{-1}
\FALabel(14.8458,15.5623)[b]{$\mu$}
\FAProp(20.,5.)(10.,6.)(0.,){/Straight}{-1}
\FALabel(15.1542,6.56231)[b]{$q'$}
\FAProp(10.,10.)(10.,14.)(0.,){/Sine}{1}
\FALabel(11.07,12.)[l]{$W$}
\FAProp(10.,10.)(10.,6.)(0.,){/Sine}{-1}
\FALabel(8.93,8.)[r]{$W$}
\FAVert(10.,14.){0}
\FAVert(10.,6.){0}
\FAVert(10.,10.){-1}
\end{feynartspicture}
} \vspace*{-2em}
\caption{Generic diagrams for the one-loop corrections to
$\nu_\mu q \to \nu_\mu q$ and 
$\nu_\mu q \to \mu^- q'$.}
\label{fig:gendiags}
\end{figure}

\begin{figure}
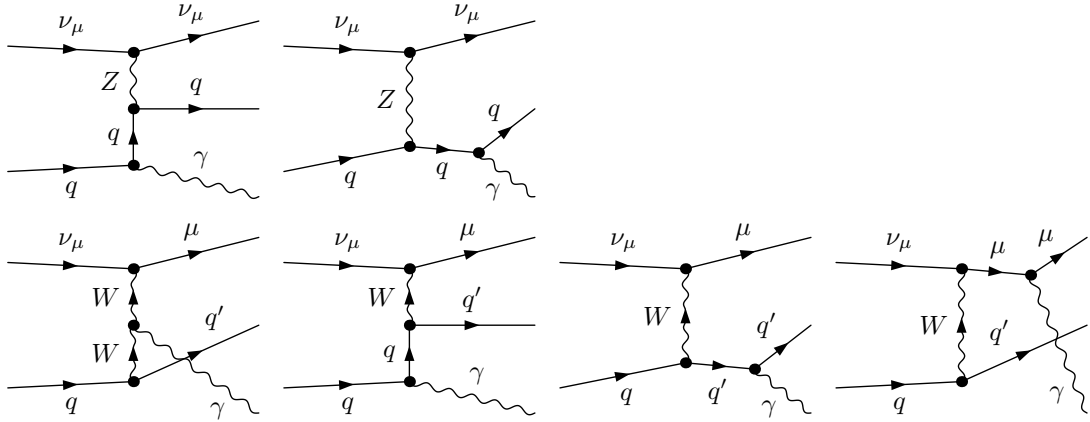

\centerline{\footnotesize  
\unitlength=1bp%
\begin{feynartspicture}(532,104)(4,1)
\FADiagram{}
\FAProp(0.,15.)(10.,14.5)(0.,){/Straight}{1}
\FALabel(5.0774,15.8181)[b]{$\nu_\mu$}
\FAProp(0.,5.)(10.,5.5)(0.,){/Straight}{1}
\FALabel(5.0774,4.18193)[t]{$q$}
\FAProp(20.,17.)(10.,14.5)(0.,){/Straight}{-1}
\FALabel(14.6241,16.7737)[b]{$\nu_\mu$}
\FAProp(20.,10.)(10.,10.)(0.,){/Straight}{-1}
\FALabel(15.,11.07)[b]{$q$}
\FAProp(20.,3.)(10.,5.5)(0.,){/Sine}{0}
\FALabel(15.3759,5.27372)[b]{$\gamma$}
\FAProp(10.,14.5)(10.,10.)(0.,){/Sine}{0}
\FALabel(8.93,12.25)[r]{$Z$}
\FAProp(10.,5.5)(10.,10.)(0.,){/Straight}{1}
\FALabel(8.93,7.75)[r]{$q$}
\FAVert(10.,14.5){0}
\FAVert(10.,5.5){0}
\FAVert(10.,10.){0}
\FADiagram{}
\FAProp(0.,15.)(10.,14.5)(0.,){/Straight}{1}
\FALabel(5.0774,15.8181)[b]{$\nu_\mu$}
\FAProp(0.,5.)(10.,7.)(0.,){/Straight}{1}
\FALabel(5.30398,4.9601)[t]{$q$}
\FAProp(20.,17.)(10.,14.5)(0.,){/Straight}{-1}
\FALabel(14.6241,16.7737)[b]{$\nu_\mu$}
\FAProp(20.,10.)(15.5,6.5)(0.,){/Straight}{-1}
\FALabel(17.2784,8.9935)[br]{$q$}
\FAProp(20.,3.)(15.5,6.5)(0.,){/Sine}{0}
\FALabel(17.2784,4.0065)[tr]{$\gamma$}
\FAProp(10.,14.5)(10.,7.)(0.,){/Sine}{0}
\FALabel(8.93,10.75)[r]{$Z$}
\FAProp(10.,7.)(15.5,6.5)(0.,){/Straight}{1}
\FALabel(12.6097,5.68637)[t]{$q$}
\FAVert(10.,14.5){0}
\FAVert(10.,7.){0}
\FAVert(15.5,6.5){0}
\FADiagram{}
\FADiagram{}
\end{feynartspicture}
} 
\vspace*{-2em}
\centerline{\footnotesize  
\unitlength=1bp%
\begin{feynartspicture}(532,104)(4,1)
\FADiagram{}
\FAProp(0.,15.)(10.,14.5)(0.,){/Straight}{1}
\FALabel(5.0774,15.8181)[b]{$\nu_\mu$}
\FAProp(0.,5.)(10.,5.5)(0.,){/Straight}{1}
\FALabel(5.0774,4.18193)[t]{$q$}
\FAProp(20.,17.)(10.,14.5)(0.,){/Straight}{-1}
\FALabel(14.6241,16.7737)[b]{$\mu$}
\FAProp(20.,10.)(10.,5.5)(0.,){/Straight}{-1}
\FALabel(17.2909,9.58086)[br]{$q'$}
\FAProp(20.,3.)(10.,10.)(0.,){/Sine}{0}
\FALabel(17.3366,3.77519)[tr]{$\gamma$}
\FAProp(10.,14.5)(10.,10.)(0.,){/Sine}{-1}
\FALabel(8.93,12.25)[r]{$W$}
\FAProp(10.,5.5)(10.,10.)(0.,){/Sine}{1}
\FALabel(8.93,7.75)[r]{$W$}
\FAVert(10.,14.5){0}
\FAVert(10.,5.5){0}
\FAVert(10.,10.){0}
\FADiagram{}
\FAProp(0.,15.)(10.,14.5)(0.,){/Straight}{1}
\FALabel(5.0774,15.8181)[b]{$\nu_\mu$}
\FAProp(0.,5.)(10.,5.5)(0.,){/Straight}{1}
\FALabel(5.0774,4.18193)[t]{$q$}
\FAProp(20.,17.)(10.,14.5)(0.,){/Straight}{-1}
\FALabel(14.6241,16.7737)[b]{$\mu$}
\FAProp(20.,10.)(10.,10.)(0.,){/Straight}{-1}
\FALabel(15.,11.07)[b]{$q'$}
\FAProp(20.,3.)(10.,5.5)(0.,){/Sine}{0}
\FALabel(15.3759,5.27372)[b]{$\gamma$}
\FAProp(10.,14.5)(10.,10.)(0.,){/Sine}{-1}
\FALabel(8.93,12.25)[r]{$W$}
\FAProp(10.,5.5)(10.,10.)(0.,){/Straight}{1}
\FALabel(8.93,7.75)[r]{$q$}
\FAVert(10.,14.5){0}
\FAVert(10.,5.5){0}
\FAVert(10.,10.){0}
\FADiagram{}
\FAProp(0.,15.)(10.,14.5)(0.,){/Straight}{1}
\FALabel(5.0774,15.8181)[b]{$\nu_\mu$}
\FAProp(0.,5.)(10.,7.)(0.,){/Straight}{1}
\FALabel(5.30398,4.9601)[t]{$q$}
\FAProp(20.,17.)(10.,14.5)(0.,){/Straight}{-1}
\FALabel(14.6241,16.7737)[b]{$\mu$}
\FAProp(20.,10.)(15.5,6.5)(0.,){/Straight}{-1}
\FALabel(17.2784,8.9935)[br]{$q'$}
\FAProp(20.,3.)(15.5,6.5)(0.,){/Sine}{0}
\FALabel(17.2784,4.0065)[tr]{$\gamma$}
\FAProp(10.,14.5)(10.,7.)(0.,){/Sine}{-1}
\FALabel(8.93,10.75)[r]{$W$}
\FAProp(10.,7.)(15.5,6.5)(0.,){/Straight}{1}
\FALabel(12.6097,5.68637)[t]{$q'$}
\FAVert(10.,14.5){0}
\FAVert(10.,7.){0}
\FAVert(15.5,6.5){0}
\FADiagram{}
\FAProp(0.,15.)(10.,14.5)(0.,){/Straight}{1}
\FALabel(5.0774,15.8181)[b]{$\nu_\mu$}
\FAProp(0.,5.)(10.,5.5)(0.,){/Straight}{1}
\FALabel(5.0774,4.18193)[t]{$q$}
\FAProp(20.,17.)(15.5,14.)(0.,){/Straight}{-1}
\FALabel(17.3702,16.3097)[br]{$\mu$}
\FAProp(20.,10.)(10.,5.5)(0.,){/Straight}{-1}
\FALabel(13.8364,8.43358)[br]{$q'$}
\FAProp(20.,3.)(15.5,14.)(0.,){/Sine}{0}
\FALabel(18.2486,4.40534)[r]{$\gamma$}
\FAProp(10.,14.5)(10.,5.5)(0.,){/Sine}{-1}
\FALabel(8.93,10.)[r]{$W$}
\FAProp(10.,14.5)(15.5,14.)(0.,){/Straight}{1}
\FALabel(12.8903,15.3136)[b]{$\mu$}
\FAVert(10.,14.5){0}
\FAVert(10.,5.5){0}
\FAVert(15.5,14.){0}
\end{feynartspicture}
} \vspace*{-2em}
\caption{Bremsstrahlung diagrams for the processes
$\nu_\mu q \to \nu_\mu q$ and 
$\nu_\mu q \to \mu^- q'$.}
\label{fig:bremdiags}
\end{figure}

We have calculated the cross sections to the processes \refeq{eq:ncprocesses}
and \refeq{eq:ccprocesses}
including relative ${\cal O}(\alpha)$ electroweak corrections. 
This means our partonic cross sections $\sigma$ can be written as a sum of 
leading-order ($\sigma_0$), virtual ($\sigma_\virt$), 
and real ($\sigma_\real$) contributions: 
\begin{equation}
\label{eq:sigma}
\sigma = \sigma_0 + \sigma_\virt + \sigma_\real 
= \sigma_0(1+\delta),
\end{equation}
where $\de$ denotes the relative correction. The virtual correction
$\sigma_\virt$ contains 
one-loop diagrams and counterterms,
\begin{equation}
\label{eq:sigvirt}
\sigma_\virt = \sigma_\onel + \sigma_\ct.
\end{equation}
The generic diagrams for these contributions, with loops and counterterms
summarized by the blobs, are shown in \reffi{fig:gendiags}.
Within the real correction $\sigma_\real$ we can discriminate 
bremsstrahlung corrections ($\sigma_\brem$)
and a counterterm contribution ($\sigma_\coll$) that accounts for 
the absorption of initial-state mass singularities into
the parton densities,
\begin{equation}
\label{eq:sigreal}
\sigma_\real = \sigma_\brem + \sigma_\coll.
\label{eq:sigma_real}
\end{equation}
The bremsstrahlung diagrams are shown in \reffi{fig:bremdiags}.

The contributions to $\sigma_\virt$ and $\sigma_\real$ diverge 
individually and cannot be evaluated unless
suitable regularization and renormalization prescriptions are employed.
All parts of our calculation have been performed in two
independent ways, resulting in two completely independent computer
codes. Both loop calculations are carried out in `t~Hooft--Feynman gauge
and are based on the standard techniques for
one-loop integrations as, e.g., described in 
\citeres{'tHooft:1979xw,Denner:1993kt}. Ultraviolet divergences
are treated in dimensional regularization and eliminated using the
on-shell renormalization scheme \cite{Denner:1993kt,Bohm:rj} in the
formulation of \citere{Denner:1993kt}.
Infrared (i.e.\ soft and collinear) singularities are regularized
by an infinitesimal photon mass and small fermion masses.
The artificial photon-mass dependence of the virtual and (soft) real 
corrections cancels in the sum of both contributions, according to
Bloch and Nordsieck \cite{Bloch:1937pw}. The role of the other fermion-mass
singularities will be discussed below.

In detail, the two calculations of the ${\cal O}(\alpha)$ corrections
have been performed as follows:
\begin{enumerate}
\item
In the first calculation,
the fully differential partonic $2 \to 2,3$ particle
cross sections have been obtained
with the Mathematica packages {\sc FeynArts} \cite{Kublbeck:1990xc} and
{\sc FormCalc} \cite{Hahn:1998yk}. 
The one-loop integrals are evaluated with the 
{\sc LoopTools} library \cite{Hahn:1998yk}.
The combination of virtual and real corrections is done by
phase-space slicing in the soft-photon region, i.e.\
soft photons are excluded from the phase-space integral of the
bremsstrahlung by a small energy cutoff $\Delta E < E_\gamma$
and added as correction factor to the lowest-order cross section
derived from the eikonal current (see e.g.\ 
\citeres{Denner:1993kt,Bloch:1937pw}).
In the whole calculation, fermion masses are kept small but non-zero
in the matrix elements and in the phase space.
\item
The second calculation is completely independent of the first.
The results for the CC processes \refeq{eq:ccprocesses} are 
obtained via crossing the results of \citere{Dittmaier:2001ay} where the 
related Drell--Yan-like process $q\bar q'\to l^-\bar\nu_l$ was treated in
${\cal O}(\alpha)$. Of course, the W-boson width, needed there to
describe the W-boson resonance, is set to zero in our case.
The corrections to the NC processes \refeq{eq:ncprocesses}
have been calculated in the same way as described in 
\citere{Dittmaier:2001ay} for the CC case.
This means that the one-loop amplitudes are generated with  {\sc FeynArts} 
\cite{Kublbeck:1990xc} and algebraically reduced using {\sc FeynCalc}
\cite{Mertig:1991an} or own independent 
{\sc Mathematica} routines; 
the scalar and tensor one-loop integrals are
evaluated using the methods and results of 
\citeres{'tHooft:1979xw,Denner:1993kt}. 
The bremsstrahlung amplitudes are derived in the Weyl--van der Waerden
spinor technique, following the formulation of \citere{Dittmaier:1999nn}.
The combination of virtual and real corrections is performed in three
different ways: using two different versions of phase-space slicing,
which are described in \citere{Dittmaier:2001ay} in detail, and
the dipole subtraction approach, as described in \citere{Dittmaier:2000mb}.
Using these techniques, masses of the external fermions can be set to 
zero in all bremsstrahlung
amplitudes consistently; the mass dependence relevant for
mass-singular regions is restored according to factorization properties
of amplitudes.
\end{enumerate}
We have checked that the different calculations lead to results on the
relative ${\cal O}(\alpha)$ corrections to the NC and CC cross sections
that typically agree within $\sim 0.1\%$. 
For the ${\cal O}(\alpha)$-induced shift $\De\sin^2\theta_\PW$ in the
weak mixing angle (see \refse{sec:Rnu} below) we find agreement within
$\sim 0.0003$.
The residual differences, which are due to the 
various ways of treating the fermion-mass effects, are thus below
the NuTeV theoretical uncertainties related to other effects
\cite{Zeller:2001hh,McFarland:2003jw}.

In the following, we do not present the full formulas for all contributions
to the cross section%
\footnote{For the CC processes all contributions can be obtained from the
formulas of \citere{Dittmaier:2001ay} via crossing.}, 
since their derivation is by now a standard exercise
and their form does not reveal further insight. We restrict ourselves to
sketching salient features and delicate parts.

\subsection{Contributions from 2-particle final states --
tree level and virtual corrections}
\label{sec:2to2}

The differential cross section including all $2\to2$ particle contributions
at ${\cal O}(\alpha^3)$ is given by 
\beq
\frac{\rd\sigma}{\rd\cos\theta}
= \frac{1}{2} \, \frac{1}{32\pi s} \,
\sum_{\tau=\pm} 
\left(|\M^\tau_0|^2 + 2\Re\{(\M_0^\tau)^* \M_1^\tau\}\right),
\eeq
where $\M_0$ and $\M_1$ are the amplitudes at tree and one-loop level,
respectively. Note that rotational invariance has been used to integrate
over the azimuthal angle $\phi$, leading to a global factor $2\pi$,
and the explicit prefactor $1/2$ accounts for the spin average of the
initial-state quark.

\begin{figure}
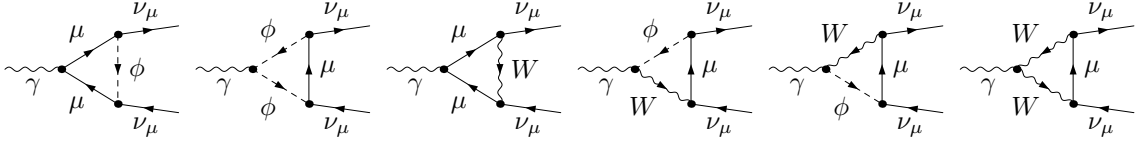

\centerline{\footnotesize  
\unitlength=1bp%
\begin{feynartspicture}(432,84)(6,1)
\FADiagram{}
\FAProp(0.,10.)(6.5,10.)(0.,){/Sine}{0}
\FALabel(3.25,8.93)[t]{$\gamma$}
\FAProp(20.,15.)(13.,14.)(0.,){/Straight}{-1}
\FALabel(16.2808,15.5544)[b]{$\nu_\mu$}
\FAProp(20.,5.)(13.,6.)(0.,){/Straight}{1}
\FALabel(16.2808,4.44558)[t]{$\nu_\mu$}
\FAProp(6.5,10.)(13.,14.)(0.,){/Straight}{1}
\FALabel(9.20801,13.1807)[br]{$\mu$}
\FAProp(6.5,10.)(13.,6.)(0.,){/Straight}{-1}
\FALabel(9.20801,6.81927)[tr]{$\mu$}
\FAProp(13.,14.)(13.,6.)(0.,){/ScalarDash}{1}
\FALabel(14.274,10.)[l]{$\phi$}
\FAVert(6.5,10.){0}
\FAVert(13.,14.){0}
\FAVert(13.,6.){0}
\FADiagram{}
\FAProp(0.,10.)(6.5,10.)(0.,){/Sine}{0}
\FALabel(3.25,8.93)[t]{$\gamma$}
\FAProp(20.,15.)(13.,14.)(0.,){/Straight}{-1}
\FALabel(16.2808,15.5544)[b]{$\nu_\mu$}
\FAProp(20.,5.)(13.,6.)(0.,){/Straight}{1}
\FALabel(16.2808,4.44558)[t]{$\nu_\mu$}
\FAProp(6.5,10.)(13.,14.)(0.,){/ScalarDash}{-1}
\FALabel(9.20801,13.1807)[br]{$\phi$}
\FAProp(6.5,10.)(13.,6.)(0.,){/ScalarDash}{1}
\FALabel(9.20801,6.81927)[tr]{$\phi$}
\FAProp(13.,14.)(13.,6.)(0.,){/Straight}{-1}
\FALabel(14.274,10.)[l]{$\mu$}
\FAVert(6.5,10.){0}
\FAVert(13.,14.){0}
\FAVert(13.,6.){0}
\FADiagram{}
\FAProp(0.,10.)(6.5,10.)(0.,){/Sine}{0}
\FALabel(3.25,8.93)[t]{$\gamma$}
\FAProp(20.,15.)(13.,14.)(0.,){/Straight}{-1}
\FALabel(16.2808,15.5544)[b]{$\nu_\mu$}
\FAProp(20.,5.)(13.,6.)(0.,){/Straight}{1}
\FALabel(16.2808,4.44558)[t]{$\nu_\mu$}
\FAProp(6.5,10.)(13.,14.)(0.,){/Straight}{1}
\FALabel(9.20801,13.1807)[br]{$\mu$}
\FAProp(6.5,10.)(13.,6.)(0.,){/Straight}{-1}
\FALabel(9.20801,6.81927)[tr]{$\mu$}
\FAProp(13.,14.)(13.,6.)(0.,){/Sine}{1}
\FALabel(14.274,10.)[l]{$W$}
\FAVert(6.5,10.){0}
\FAVert(13.,14.){0}
\FAVert(13.,6.){0}
\FADiagram{}
\FAProp(0.,10.)(6.5,10.)(0.,){/Sine}{0}
\FALabel(3.25,8.93)[t]{$\gamma$}
\FAProp(20.,15.)(13.,14.)(0.,){/Straight}{-1}
\FALabel(16.2808,15.5544)[b]{$\nu_\mu$}
\FAProp(20.,5.)(13.,6.)(0.,){/Straight}{1}
\FALabel(16.2808,4.44558)[t]{$\nu_\mu$}
\FAProp(6.5,10.)(13.,14.)(0.,){/ScalarDash}{-1}
\FALabel(9.20801,13.1807)[br]{$\phi$}
\FAProp(6.5,10.)(13.,6.)(0.,){/Sine}{1}
\FALabel(9.20801,6.81927)[tr]{$W$}
\FAProp(13.,14.)(13.,6.)(0.,){/Straight}{-1}
\FALabel(14.274,10.)[l]{$\mu$}
\FAVert(6.5,10.){0}
\FAVert(13.,14.){0}
\FAVert(13.,6.){0}
\FADiagram{}
\FAProp(0.,10.)(6.5,10.)(0.,){/Sine}{0}
\FALabel(3.25,8.93)[t]{$\gamma$}
\FAProp(20.,15.)(13.,14.)(0.,){/Straight}{-1}
\FALabel(16.2808,15.5544)[b]{$\nu_\mu$}
\FAProp(20.,5.)(13.,6.)(0.,){/Straight}{1}
\FALabel(16.2808,4.44558)[t]{$\nu_\mu$}
\FAProp(6.5,10.)(13.,14.)(0.,){/Sine}{-1}
\FALabel(9.20801,13.1807)[br]{$W$}
\FAProp(6.5,10.)(13.,6.)(0.,){/ScalarDash}{1}
\FALabel(9.20801,6.81927)[tr]{$\phi$}
\FAProp(13.,14.)(13.,6.)(0.,){/Straight}{-1}
\FALabel(14.274,10.)[l]{$\mu$}
\FAVert(6.5,10.){0}
\FAVert(13.,14.){0}
\FAVert(13.,6.){0}
\FADiagram{}
\FAProp(0.,10.)(6.5,10.)(0.,){/Sine}{0}
\FALabel(3.25,8.93)[t]{$\gamma$}
\FAProp(20.,15.)(13.,14.)(0.,){/Straight}{-1}
\FALabel(16.2808,15.5544)[b]{$\nu_\mu$}
\FAProp(20.,5.)(13.,6.)(0.,){/Straight}{1}
\FALabel(16.2808,4.44558)[t]{$\nu_\mu$}
\FAProp(6.5,10.)(13.,14.)(0.,){/Sine}{-1}
\FALabel(9.20801,13.1807)[br]{$W$}
\FAProp(6.5,10.)(13.,6.)(0.,){/Sine}{1}
\FALabel(9.20801,6.81927)[tr]{$W$}
\FAProp(13.,14.)(13.,6.)(0.,){/Straight}{-1}
\FALabel(14.274,10.)[l]{$\mu$}
\FAVert(6.5,10.){0}
\FAVert(13.,14.){0}
\FAVert(13.,6.){0}
\end{feynartspicture}
} 
 \vspace*{-2em}
\caption{Diagrams for the one-loop correction to the 
$\gamma\nu_\mu\bar\nu_\mu$ vertex.}
\label{fig:anndiags}
\end{figure}

Among the virtual corrections, there are two contributions to the
NC processes that become numerically delicate in the limit of small
momentum transfer ($t\to0$):
the $\gamma\nu_\mu\bar\nu_\mu$ vertex correction and the $\gamma Z$
mixing self-energy, indicated in 
the last two generic diagrams in the first line of \reffi{fig:gendiags}.
The corresponding contributions to the relative correction $\de$ read
\beqar
\de^\tau_{\gamma\nu_\mu\bar\nu_\mu} &=&
-\frac{4\sw\cw Q_q}{g^\tau_{qqZ}} \frac{t-\MZ^2}{t}
\left[ F_{\gamma\nu_\mu\bar\nu_\mu}(t) 
+\frac{1}{4\sw\cw}\de Z_{Z\gamma}\right],
\label{eq:nna}
\\
\de_{\gamma Z}^\tau &=&
\frac{2Q_q}{g^\tau_{qqZ}} \left[
\frac{\Sigma^{\gamma Z}_{\mathrm{T}}(t)}{t}
+\frac{1}{2}\de Z_{\gamma Z}+\frac{t-\MZ^2}{2t}\de Z_{Z\gamma} \right] \label{eq:gZ},
\eeqar
where $\Sigma^{\gamma Z}_{\mathrm{T}}$ is the transversal part 
of the unrenormalized $\gamma Z$ mixing self-energy and
$\de Z_{\gamma Z}$ and $\de Z_{Z\gamma}$ are 
the renormalization constants for the mixing of the $\gamma$ and 
$Z$ fields.
The precise definition and explicit expressions for these
quantities in `t~Hooft--Feynman gauge
can be found in \citere{Denner:1993kt}.
The one-loop diagrams contributing to the $\gamma\nu_\mu\bar\nu_\mu$ vertex
are depicted in \reffi{fig:anndiags}, where $\phi$ represents the
would-be Goldstone field corresponding to the W boson.
The renormalized vertex form factor 
$F_{\gamma\nu_\mu\bar\nu_\mu}^{\mathrm{ren}}(t)$, which is given by the
expression in square brackets in Eq.~\refeq{eq:nna},
explicitly reads
\beqar
F_{\gamma\nu_\mu\bar\nu_\mu}^{\mathrm{ren}}(t) &=&
       -\frac{\alpha}{16\pi\sw^2} \biggl\{
       \frac{2(m_\mu^2+2 \MW^2)}{\MW^2}
       -4[B_0(0,\MW,\MW)-B_0(t,\MW,\MW)]
\nn\\ && {}
       -\frac{2m_\mu^4+2m_\mu^2\MW^2-4\MW^4+m_\mu^2 t-6\MW^2 t}{t\MW^2}
\nn\\ && {}
              \times [B_0(t,m_\mu,m_\mu)-B_0(t,\MW,\MW)]
\nn\\ && {}
       +\frac{2(m_\mu^6-3m_\mu^2\MW^4+2\MW^6+m_\mu^4 t
                      -2m_\mu^2\MW^2 t+4\MW^4 t)}{t\MW^2}
\nn\\ && {}
               \times [C_0(0,t,0,m_\mu,\MW,\MW) +C_0(0,t,0,\MW,m_\mu,m_\mu)] 
\nn\\ && {}
       +4 t C_0(0,t,0,\MW,m_\mu,m_\mu)
\biggr\}.
\label{eq:Fann}
\eeqar
The definitions and general results for the scalar 2- and 3-point integrals
$B_0$ and $C_0$ can be found in \citere{Denner:1993kt}.
Note that the full muon mass dependence is kept in this formula.
For large $|t|$, i.e.\ for $|t|\gg m_\mu^2$,
the muon mass can be neglected; in this limit the form factor reads
\beqar
F_{\gamma\nu_\mu\bar\nu_\mu}^{\mathrm{ren}}(t)\Big|_{m_\mu=0} &=&
       -\frac{\alpha}{16\pi\sw^2} \biggl\{
       4-4[B_0(0,\MW,\MW)-B_0(t,\MW,\MW)]
\nn\\ && {}
       +\frac{2(2\MW^2+3t)}{t} [B_0(t,0,0)-B_0(t,\MW,\MW)]
\nn\\ && {}
       +\frac{4\MW^2(\MW^2+2t)}{t}
               [C_0(0,t,0,0,\MW,\MW) +C_0(0,t,0,\MW,0,0)] 
\nn\\ && {}
       +4 t C_0(0,t,0,\MW,0,0)
\biggr\},
\label{eq:Fannmu0}
\eeqar
which reproduces the full expression for $F_{\gamma\nu_\mu\bar\nu_\mu}$
up to terms of ${\cal O}(m_\mu^2/t)$ and ${\cal O}(m_\mu^2/\MW^2)$.
For small $|t|$, the general expression \refeq{eq:Fann} becomes
numerically unstable,
because the whole expression in this limit is of ${\cal O}(t)$
while individual terms involve constants and even $1/t$ poles.
A stable result for this limit can be obtained
by performing an asymptotic expansion for large $\MW$; this leads to
\beqar
F_{\gamma\nu_\mu\bar\nu_\mu}^{\mathrm{ren}}(t) &\asymp{\MW\to\infty}&
-\frac{\alpha}{36\pi\MW^2\sw^2} \biggl\{
6m_\mu^2[B_0(0,m_\mu,m_\mu)-B_0(t,m_\mu,m_\mu)]
\nn\\ && {}
-t[2-3B_0(0,\MW,\MW)+3B_0(t,m_\mu,m_\mu)] \biggr\},
\label{eq:FannMW}
\eeqar
which is valid up to terms of ${\cal O}(m_\mu^4/\MW^4)$,
${\cal O}(m_\mu^2t/\MW^4)$, and ${\cal O}(t^2/\MW^4)$.
From this expression one can easily obtain the limit $t\to0$ for the
contribution of the $\gamma\nu_\mu\bar\nu_\mu$ vertex to the relative
correction,
\beqar
\de^\tau_{\gamma\nu_\mu\bar\nu_\mu}\Big|_{t=0} &=&
\frac{Q_q}{g^\tau_{qqZ}} \frac{\alpha}{3\pi\cw\sw} \biggl\{
1+\ln\biggl(\frac{\MW^2}{m_\mu^2}\biggr) \biggr\}.
\eeqar

In the limit $t\to0$, the relative correction induced by $\gamma Z$ mixing 
tends to the value
\beqar
\de_{\gamma Z}^\tau\Big|_{t=0} &=&
\frac{2Q_q}{g^\tau_{qqZ}} \left[
\frac{\Sigma^{\gamma Z}_{\mathrm{T}}(0)
-\Sigma^{\gamma Z}_{\mathrm{T}}(\MZ^2) }{\MZ^2}
+ \Sigma^{\gamma Z\prime }_{\mathrm{T}}(0)
\right],
\label{eq:deAZ0}
\eeqar
where $\Sigma^{\gamma Z\prime }_{\mathrm{T}}(t)$ is the derivative of
$\Sigma^{\gamma Z}_{\mathrm{T}}(t)$ w.r.t.\ the variable $t$.
Since the derivative $\Sigma^{\gamma Z\prime }_{\mathrm{T}}(0)$ 
numerically results from 
$[\Sigma^{\gamma Z}_{\mathrm{T}}(t)
-\Sigma^{\gamma Z}_{\mathrm{T}}(0)]/t$
for $t\to0$,
the actual evaluation of $\de_{\gamma Z}^\tau$
deserves some care for small $|t|$. 
This is, however,
less tricky than for $\de^\tau_{\gamma\nu_\mu\bar\nu_\mu}$, since
only one power of $t$ effectively cancels for $t\to0$.
A peculiarity in $\Sigma^{\gamma Z\prime }_{\mathrm{T}}(0)$ is
that it involves logarithms $\ln(m_f/\MZ)$ of all charged fermions $f$.
These contributions result from the $\gamma Z$ mixing in the 
non-perturbative domain of small momentum transfer. 
Thus, the 
light-quark masses entering here should be tuned in such a way that the
hadronic $\gamma Z$ mixing, which can be obtained via dispersion relations
\cite{Jegerlehner:1985gq}, is reproduced.
This highly non-trivial task is, however, beyond the scope of this paper. 

In the actual numerical evaluation, 
we proceed in the two following ways
(corresponding to the two independent calculations described under the same
labels in \refse{se:calcframe}):
\begin{enumerate}
\item
In the first calculation we substitute 
$\de^\tau_{\gamma\nu_\mu\bar\nu_\mu}\Big|_{t=0}$ and
$\de_{\gamma Z}^\tau\Big|_{t=0}$ for the exact $t$-dependent expression if
$|t| < 5\times 10^{-3}\GeV^2$.
\item
In the second calculation the asymptotic formulas 
\refeq{eq:Fannmu0} or \refeq{eq:FannMW} are used if $|t|$ is larger or smaller
than $1\GeV^2$, respectively. At the matching point, these two formulas
agree within about 4 digits. The $\ga Z$ mixing part is evaluated with the
full expression \refeq{eq:nna} everywhere.
\end{enumerate}

\subsection{Contributions from 3-particle final states -- bremsstrahlung 
corrections}
\label{sec:2to3}

For the real photonic corrections, the processes \refeq{eq:ncprocesses}
and \refeq{eq:ccprocesses} have to be considered with an additional
photon (with momentum $k$) in the final state.
The contribution $\sigma_\brem$ of these radiative processes to the
parton cross section are given by
\beq
\sigma_\brem = \frac{1}{2}\frac{1}{2\hat s} \int \rd\Gamma_\gamma \,
\sum_{\rm spins} |\M_\gamma|^2,
\label{eq:hbcs}
\eeq
where $\M_\gamma$ is the corresponding amplitude and
the phase-space integral is defined by
\beq
\int \rd\Gamma_\gamma =
\int\frac{\rd^3 {\bf k}_l}{(2\pi)^3 2k_{l,0}}
\int\frac{\rd^3 {\bf k}_q}{(2\pi)^3 2k_{q,0}}
\int\frac{\rd^3 {\bf k}}{(2\pi)^3 2k_0} \,
(2\pi)^4 \delta(p_l+p_q-k_l-k_q-k).
\label{eq:dGg}
\eeq
The phase-space integral \refeq{eq:hbcs} diverges in the soft ($k_0\to
0$) and collinear ($p_q k, k_q k \to 0$, and $k_l k\to0$ in the CC case) 
regions logarithmically if
the photon and fermion masses are set to zero. For the treatment of
the soft and collinear singularities we have applied the phase-space
slicing method as well as a subtraction approach, as mentioned above.

\subsection{Hadron cross section and redefinition of parton densities}
\label{sec:partdens}

The neutrino--nucleus scattering cross section $\sigma_{\nu N}$ 
is obtained from the
parton cross sections $\sigma$ by convolution with the
corresponding parton distribution functions $q(x)$,
\beq
\sigma_{\nu N}(E_\nu^{\LAB}) = \sum_q \int_0^1 \rd x \,
q_\iso(x) \,\sigma(p_l, p_q),
\label{eq:signN}
\eeq
where $x$ is the momentum fraction carried by the parton $q$ and $M$ is
the factorization scale.
The squared CM energy $s$ of the partonic process
is related to the neutrino energy $E_\nu^{\LAB}$ in the LAB frame
according to Eq.\refeq{eq:s_EnuLAB}.
In the sum $\sum_q$ the variable $q$ runs over the incoming
quarks and antiquarks indicated in Eqs.~\refeq{eq:ncprocesses}
and \refeq{eq:ccprocesses}. The subscript ``iso'' indicates that
we average over the parton densities for an (anti-)quark $q$ in the
proton and neutron, $q(x)\equiv f^\Pp_q(x)$ and $f^\Pn_q(x)$,
\beq
q_\iso(x) = \frac{1}{2}\left( f^\Pp_q(x) + f^\Pn_q(x) \right).
\eeq
Owing to isospin invariance ($f^\Pp_\Pu(x)\equiv f^\Pn_\Pd(x)$, etc.), 
this is equivalent to an isospin 
average of the parton densities of the proton,
$\Pu_\iso(x)\equiv\Pd_\iso(x)\equiv\frac{1}{2}(\Pu(x)+\Pd(x))$, etc.

The ${\cal O}(\alpha)$-corrected parton cross section
$\sigma$ contains mass singularities of the form
$\alpha\ln(m_q)$, which are due to collinear photon radiation off the
initial-state quarks.  In complete analogy to the
$\overline{\mbox{MS}}$ factorization scheme for next-to-leading order
QCD corrections, we absorb these collinear singularities into the
quark distributions.  This is achieved by replacing $q(x)$ in
\refeq{eq:signN} according to 
\beqar
q(x) &\to& q(x,M^2) 
\nn\\
&& {} 
-\int_x^1 \frac{\rd z}{z} \, q\biggl(\frac{x}{z},M^2\biggr) \,
\frac{\alpha}{2\pi} \, Q_q^2 \, \biggl\{
\ln\biggl(\frac{M^2}{m_q^2}\biggr) \Bigl[ P_{ff}(z) \Bigr]_+
-\Bigl[ P_{ff}(z) \Bigl(2\ln(1-z)+1\Bigr) \Bigr]_+
\biggr\},
\nn\\
\label{eq:factorization}
\eeqar
where $M$ is the factorization scale (see \citere{Dittmaier:2001ay}). 
This replacement defines the
same finite parts in the ${\cal O}(\alpha)$ correction as the usual 
$\overline{\mbox{MS}}$ factorization in $D$-dimensional regularization
for exactly massless partons, where the $\ln(m_q)$ terms appear as
$1/(D-4)$ poles.
In \refeq{eq:factorization}
we have regularized the
soft-photon pole by using the $[\dots]_+$
prescription. This procedure is fully equivalent to the application of
a soft-photon cutoff $\Delta E$ (see~\citere{Baur:1999kt}) where 
\begin{eqnarray}\label{eq:factorization-cut}
q(x) &\to&
 q(x,M^2) \; \Biggl[ 1 - \frac{\alpha}{\pi} \; Q_q^2 
\Biggl\{1-\ln(2\Delta E/\sqrt{\hat{s}}) -\ln^2(2\Delta E/\sqrt{\hat{s}})
\nonumber\\ 
&& \hspace*{18mm} + \left(\ln(2\Delta E/\sqrt{\hat{s}})
+\frac{3}{4}\right)\,\ln\left(\frac{M^2}{m_q^2}\right)\Biggr\}\Biggr]
\nonumber\\[2.mm]
&&{}- \int_x^{1-2\Delta E/\sqrt{\hat{s}}} \frac{d z}{z}\; 
q\left(\frac{x}{z},M^2\right)
\; \frac{\alpha}{2 \pi} \, Q_q^2 \,
P_{ff}(z)\left\{
\ln\left(\frac{M^2}{m_q^2}\frac{1}{(1-z)^2}\right)
-1\right\}.
\hspace{2em}
\label{eq:factorization2}
\end{eqnarray}
The convolution of the correction terms to the parton densities
[r.h.s.\ of Eqs.~\refeq{eq:factorization} or \refeq{eq:factorization2}]
with the lowest-order cross section $\sigma_0$ defines the ``collinear
counterterm'' contribution $\sigma_\coll$ in the real photonic
correction $\sigma_\real$, as defined in Eq.~\refeq{eq:sigma_real}.

The absorption of the collinear singularities of ${\cal O}(\alpha)$
into quark distributions, as a matter of fact, requires also the
inclusion of the corresponding ${\cal O}(\alpha)$ corrections into the
DGLAP evolution of these distributions and into their fit to
experimental data.
At present, this full incorporation of ${\cal
O}(\alpha)$ effects in the determination of the quark distributions
has not yet been performed. However, an approximate inclusion of the
${\cal O}(\alpha)$ corrections to the DGLAP evolution shows
\cite{Kripfganz:1988bd} that the impact of these corrections on the
quark distributions in the $\overline{\mbox{MS}}$ factorization scheme
is smaller than remaining QCD uncertainties;
this is also supported by a recent analysis of the MRST collaboration
\cite{Stirling} who took into account the ${\cal O}(\alpha)$ effects
to the DGLAP equations.%
\footnote{Note that the inclusion of ${\cal O}(\alpha)$ effects in
the DGLAP evolution also leads to a photon distribution function in
the nucleon. The impact of the resulting $\nu_\mu\gamma$ reactions,
which are connected to the bremsstrahlung corrections via crossing,
is, however, expected to be very small owing to smallness of the
induced photon distribution.}

\subsection{Structure of divergences in the radiative corrections}
\label{sec:div}

While ultraviolet divergences are absorbed by renormalization and
soft singularities always compensate between virtual and (soft)
real corrections, the issue of collinear singularities is more
delicate. These singularities originate from collinear photon emission off
the external fermions and lead to mass logarithms $\alpha\ln m_f$.
As described in the previous section, the $\alpha\ln m_q$ terms
from initial-state radiation are absorbed into the parton
distribution functions.

However, there are also $\alpha\ln m_{q'}$ and (in the CC case)
$\alpha\ln m_\mu$ terms from final-state radiation.
According to the Kinoshita--Lee--Nauenberg (KLN) theorem \cite{KLN},
these terms drop out if the final state is treated sufficiently inclusive,
i.e.\ if the cones for quasi-collinear photon emission around the charged
outgoing fermions are fully integrated over.
This condition is, in general, not fulfilled if phase-space cuts at the
parton level are applied. For instance, restricting the outgoing
hadronic energy,
$E_{\had,\min}^{\LAB}<E_{\had}^{\LAB}<E_{\had,\max}^{\LAB}$, 
spoils the inclusiveness w.r.t.\ photon emission collinear to the outgoing
quark, so that corrections proportional to $\alpha\ln m_{q'}$ survive.
On the other hand, a cut on the total hadronic+photonic energy,
$E_{\had+\phot,\min}^{\LAB}<E_{\had+\phot}^{\LAB}<E_{\had+\phot,\max}^{\LAB}$, 
which is equivalent to cutting the outgoing lepton energy,
destroys the inclusiveness w.r.t.\ collinear photon emission of muons
in the CC processes, leading to $\alpha\ln m_\mu$ corrections.
Both types of mass singularities are avoided if {\it photon recombination} in 
the final state is taken into account before applying selection cuts.

For instance, a simple recombination scheme is provided by the following
algorithm:
\begin{enumerate}
\item
For an event with a photon in the final state,
the smallest angle $\theta_{\ga f}$ between the outgoing photon and
the outgoing fermions ($q$ for NC; $\mu$ or $q'$ for CC) is determined in the
LAB frame. 
\item
If $\theta_{\ga f}<\theta_{\cut}$, the photon is recombined with the nearest
fermion $f$. This means that the
momenta of the photon and of the fermion $f$ are added and
assigned to the new ``quasi-fermion'' $\tilde f$; then 
the photon is discarded from the event.
\item
Finally, we impose the hadronic energy cut $E_{\had,\recomb}^{\LAB} > 10\GeV$,
where ``$\recomb$'' indicates that the momentum after possible recombination
is used.
\end{enumerate}
This procedure is ``collinear safe'' with respect to initial-state
and final-state collinear singularities, i.e.\ photons that are
nearly collinear to a charged fermion are treated inclusively.
Note that photons exactly collinear to the initial-state quark receive
zero momentum in the LAB frame, which effectively acts as
recombination with the initial state.
All final-state muon and quark-mass logarithms cancel in the corresponding
integrated cross section. The collinear safety of observables can be
ideally checked by comparing results obtained with phase-space slicing 
to the ones obtained with a subtraction technique; we have performed
this comparison and find agreement within the integration errors.
In the numerical evaluation we take $\theta_{\cut}=5^\circ$.

Although all of the above cuts are idealizations, the cut on the
sum $E_{\had+\phot}^{\LAB}$ of hadronic and photonic energies seems
to be most appropriate for a fixed-target experiment where all but the
energy of neutrinos and muons is deposited in a calorimeter.

\section{Discussion of results}
\label{se:results}

\subsection{The quantities $\delta R^\nu$ and $\Delta \sin^2 \theta_W$}
\label{sec:Rnu}

Although the electroweak ${\cal O}(\alpha)$ corrections to the NC
and CC cross sections represent the central results of this paper,
it is useful to translate them into a correction to the cross-section
ratio $R^\nu=\si_\NC^\nu/\si_\CC^\nu$ introduced in Eq.~\refeq{eq:Rnu}
and to make contact to its interpretation in terms of an effective shift
in the on-shell weak mixing angle.

Writing $\delta\sigma$ for the correction to the cross section,
the (first-order) correction $\de R^\nu$ to the ratio $R^\nu$
reads
\beq
\delta R^\nu 
= R^\nu \left( \frac{\delta\sigma^\nu_\NC}{\sigma^\nu_\NC} 
	- \frac{\delta\sigma^\nu_\CC}{\sigma^\nu_\CC} \right)
= R^\nu \left( \delta R^\nu_\NC + \delta R^\nu_\CC \right).
\eeq
Following \citere{Bardin:1986bc}, the quantities
\beq
\delta R^\nu_\NC = \frac{\delta\sigma^\nu_\NC}{\sigma^\nu_\NC},
\qquad
\delta R^\nu_\CC = -\frac{\delta\sigma^\nu_\CC}{\sigma^\nu_\CC}
\eeq
have been defined for later convenience.

In order to define an effective shift $\De\sin^2\theta_\PW$,
we quote the approximate relation \cite{LlewellynSmith:ie}
\beq
R^\nu \sim \frac{1}{2}-\sin^2\theta_\PW+\frac{20}{27}\sin^4\theta_\PW,
\eeq
which results from the lowest-order neutrino deep-inelastic scattering
cross section ratio for
isoscalar targets upon neglecting the momentum transfers in the W- and Z-boson
propagators.
The correction $\delta R^\nu$ can be interpreted as a correction
$\de\sin^2\theta_\PW$ in the variable $\sin^2\theta_\PW$,
\beq
\de R^\nu \sim 
\de\left( \frac{1}{2}-\sin^2\theta_\PW+\frac{20}{27}\sin^4\theta_\PW\right)
= \left( -1+\frac{40}{27}\sin^2\theta_\PW\right) \de\sin^2\theta_\PW.
\eeq
As done in \citere{Bardin:1986bc}, we define the shift
$\De\sin^2\theta_\PW$ as the negative of the correction
$\de\sin^2\theta_\PW$. Combining the above relations, we are, thus, lead to
\beqar
\De\sin^2\theta_\PW \equiv
-\de\sin^2\theta_\PW &\equiv&
\frac{\frac{1}{2}-\sin^2\theta_\PW+\frac{20}{27}\sin^4\theta_\PW}
{1-\frac{40}{27}\sin^2\theta_\PW}
\left( \frac{\delta\sigma^\nu_\NC}{\sigma^\nu_\NC} 
	- \frac{\delta\sigma^\nu_\CC}{\sigma^\nu_\CC} \right)
\nn\\
&=& \frac{\frac{1}{2}-\sin^2\theta_\PW+\frac{20}{27}\sin^4\theta_\PW}
{1-\frac{40}{27}\sin^2\theta_\PW}
\left( \delta R^\nu_\NC + \delta R^\nu_\CC \right).
\eeqar

\subsection{Input parameters}

If not stated otherwise, we used the following set of input
parameters~\cite{Hagiwara:pw} for the numerical evaluation,
\beq
\arraycolsep 2pt
\begin{array}[b]{rclrclrcl}
\GF & = & 1.16639 \times 10^{-5} \GeV^{-2}, \quad&
\alpha(0) &=& 1/137.03599976,\quad & \alpha(M_Z) & = & 1/128.930, \\
M_N & = & m_\Pp = 0.938271998\GeV, \\
\MW & = & 80.423\GeV, &
\MZ & = & 91.1876\GeV, & 
\MH & = & 115\GeV, \\
\Me & = & 0.510998902\MeV, &
m_\mu &=& 105.658357\MeV, &
m_\tau &=& 1.77699\GeV, \\
\Mu & = & 66\MeV, &
\Mc & = & 1.2\GeV, &
\Mt & = & 174.3\;\GeV, \\
\Md & = & 66\MeV, &
\Ms & = & 150\MeV, &
\Mb & = & 4.3\GeV.
\end{array}
\label{eq:SMpar}
\eeq
The masses of the light quarks are adjusted such as to reproduce the
hadronic contribution to the photonic vacuum polarization
of~\citere{Jegerlehner:2001ca}. They are relevant only for the
evaluation of the charge renormalization constant in the
$\alpha(0)$-scheme and for the small correction induced by the
$\ga Z$ mixing at zero-momentum transfer [see Eq.~\refeq{eq:deAZ0}]. 
For the calculation of the 
$\nu_\mu N$ cross sections we have set the energy of the incoming
neutrino to $E_\nu^{\LAB}=80\GeV$ and adopted the CTEQ4L
parton distribution functions \cite{Lai:1997mg}.
The factorization scale $M$ is derived from the leptonic momentum
transfer, i.e.\ it is set to $M=\sqrt{-(p_l-k_l)^2}$
(with the momenta after possible photon recombination).

\begin{sloppypar}
For the explicit evaluation and the following numerical discussion
we distinguish three different input-parameter schemes
\begin{itemize}
\item
{\it $\alpha(0)$-scheme:}
The fine-structure constant $\alpha(0)$ and all particle masses define
the complete input (i.e.\ $\alpha(\MZ)$ and $\GF$ are not used).
In this scheme, the relative corrections to the cross sections
sensitively depend on the light quark masses via $\alpha\ln m_q$ terms
that enter the charge renormalization.
\item
{\it $\alpha(\MZ)$-scheme:}
The effective electromagnetic coupling  $\alpha(\MZ)$ 
and all particle masses define
the complete input (i.e.\ $\alpha(0)$ and $\GF$ are not used).
Tree-level couplings are derived from $\alpha(\MZ)$,
and the relative corrections are related by
$\de_{\alpha(\MZ)\mathrm{-scheme}}=
\de_{\alpha(0)\mathrm{-scheme}}-2\De\alpha(\MZ)$,
where $\De\alpha(Q)$ accounts for the running of the electromagnetic coupling
from scale $Q=0$ to $Q=\MZ$ (induced by light fermions) and cancels the
corresponding $\alpha\ln m_q$ terms in $\de_{\alpha(0)\mathrm{-scheme}}$.
\item
{\it $\GF$-scheme:}
The Fermi constant $\GF$ and all particle masses define
the complete input (i.e.\ $\alpha(0)$ and $\alpha(\MZ)$ are not used).
Tree-level couplings are derived from the effective coupling
$\alpha_{\GF}=\sqrt{2}\GF\MW^2\sw^2/\pi$, and the relative corrections
are related by
$\de_{\GF\mathrm{-scheme}}=\de_{\alpha(0)\mathrm{-scheme}}-2\De r$,
where $\De r$ contains the radiative corrections to muon decay and is
defined \cite{Sirlin:1980nh} by
\beq
\MW^2 \left(1-\frac{\MW^2}{\MZ^2}\right) = 
\frac{\pi\alpha}{\sqrt{2}\GF}\frac{1}{1-\Delta r(\alpha,\MW,\MZ,\MH,m_f)}.
\label{eq:deltar}
\eeq
In the following we consistently use $\De r$ at the one-loop level,
as, for instance, explicitly given in \citere{Denner:1993kt}.
Since $\De\alpha(\MZ)$ is contained in $\De r$, there is no large
effect induced by the running of the electromagnetic coupling in the
$\GF$-scheme either.
\end{itemize}
More details about the various schemes can, e.g., be found in
\citeres{Denner:1993kt,Bohm:rj,Dittmaier:2001ay}.
\end{sloppypar}

\subsection{Numerical results}

Using the input defined in the previous section, we have evaluated
the corrections $\delta R^\nu_\NC$ and $\delta R^\nu_\CC$, as well as
the corresponding shift $\De\sin^2\theta_\PW$, in the various 
input-parameter schemes and for different treatments of photons in the
final state.
The corresponding numerical results are collected in \refta{tab:numres}.
\begin{table}
\centerline{
$\begin{array}{ccccc}
\multicolumn{5}{l}{\mbox{Hadronic energy cut:} \quad
E_{\had}^{\LAB} > 10\GeV}
\\ 
\hline 
\mbox{IPS} & R^\nu_0 & \delta R^\nu_\NC & \delta R^\nu_\CC & \De\sin^2\theta_\PW
\\
\hline 
\alpha(0)   & 0.31766(2) & 0.0582(1)  & -0.0758(4)  & -0.0082(2)
\\
\alpha(\MZ) & 0.31766(2) & -0.0639(1) & 0.0452(4) & -0.0088(2)
\\
\GF         & 0.31766(2) & 0.0003(1)  & -0.0185(4)  & -0.0085(2)
\\
\hline 
\end{array}$ }
\vspace{1em}
\centerline{
$\begin{array}{ccccc}
\multicolumn{5}{l}{\mbox{Hadronic+photonic energy cut:} \quad
E_{\had+\phot}^{\LAB} > 10\GeV}
\\ 
\hline 
\mbox{IPS} & R^\nu_0 & \delta R^\nu_\NC & \delta R^\nu_\CC & \De\sin^2\theta_\PW
\\
\hline 
\alpha(0)   & 0.31766(2) & 0.0589(1)  & -0.0842(4)  & -0.0118(2)
\\
\alpha(\MZ) & 0.31766(2) & -0.0632(1) & 0.0363(4) & -0.0126(2)
\\
\GF         & 0.31766(2) & 0.0011(1)  & -0.0272(4)  & -0.0122(2)
\\
\hline 
\end{array}$ }
\vspace{1em}
\centerline{
$\begin{array}{ccccc}
\multicolumn{5}{l}{\mbox{Hadronic energy cut after $\ga$ recombination:} \quad
E_{\had,\recomb}^{\LAB} > 10\GeV}
\\ 
\hline 
\mbox{IPS} & R^\nu_0 & \delta R^\nu_\NC & \delta R^\nu_\CC & \De\sin^2\theta_\PW
\\
\hline 
\alpha(0)   & 0.31766(2) & 0.05861(1)  & -0.07702(2) & -0.00863(1)
\\
\alpha(\MZ) & 0.31766(2) & -0.06349(1) &  0.04392(2) & -0.00917(1)
\\
\GF         & 0.31766(2) & 0.00077(2)  & -0.01980(2) & -0.00892(1)
\\
\hline 
\end{array}$ }
\caption{Results on the ratio $R^\nu$ in lowest order ($R^\nu_0$),
corrections from NC and CC cross sections 
($\delta R^\nu_\NC$ and $\delta R^\nu_\CC$),
and shift $\De\sin^2\theta_\PW$, for various input-parameter schemes (IPS)
and using different treatments of real photons in the final state.
The number in parentheses indicates the statistical integration error
in the last digit.}
\label{tab:numres}
\end{table}

We observe a large scheme dependence of the corrections
$\delta R^\nu_\NC$ and $\delta R^\nu_\CC$ owing to the shifts of
$-2\De\alpha(\MZ)\sim-12\%$ and $-2\De r\sim-6\%$ in the relative corrections
$\de_{\alpha(\MZ)\mathrm{-scheme}}$ and $\de_{\GF\mathrm{-scheme}}$,
respectively, relative to the $\alpha(0)$-scheme. Since these shifts
are identical in the NC and CC cross sections, they drop out in 
$\De\sin^2\theta_\PW$. The remaining very small scheme dependence in
$\De\sin^2\theta_\PW$ is due to the different choice of the coupling $\alpha$
in the schemes, i.e.\ of formal two-loop order.

Moreover, \refta{tab:numres} reveals that different ways of treating
real photons in the final state selection lead to sizeable differences
in the correction $\delta R^\nu_\CC$ to the CC cross section while
$\delta R^\nu_\NC$ hardly changes. The differences can be explained
by inspecting final-state radiation w.r.t.\ their inclusiveness.
The variant $E_{\had+\phot}^{\LAB} > 10\GeV$, where the energy cut
is applied to the sum of hadronic and photonic energy, is the only
choice in which the collinear cone for radiation off outgoing muons
is not treated inclusively, leading to logarithmically enhanced
corrections proportional to $\alpha\ln m_\mu$. Numerically this
effect changes $\delta R^\nu_\CC$ by $\sim 0.008$, and thus increases
the shift $\De\sin^2\theta_\PW$ by $\sim 0.004$. 
On the other hand, the variant $E_{\had}^{\LAB} > 10\GeV$, where
the energy cut is applied to the bare outgoing quark energy, is
the only case in which final-state quark logarithms 
$\alpha\ln m_q$ (in the NC case) or $\alpha\ln m_{q'}$ (in the CC case)
are present. Numerically these effects turn out to be of the order
$\sim 0.001$ in $\delta R^\nu_\NC$ and $\delta R^\nu_\CC$, and 
thus do not influence $\De\sin^2\theta_\PW$ very much.

\subsection{Comparison with existing results}
\label{se:compBD}

Finally, we compare our results with the ones of the older
calculation of \citere{Bardin:1986bc} as far as possible.
To this end, we adopt the (obsolete) input parameters of \citere{Bardin:1986bc}
which are, however, unfortunately not completely specified there.%
\footnote{The full set of input parameters used in \citere{Bardin:1986bc}
could also not be provided by the authors on request.}
The following parameters are explicitly given:
\beq
\arraycolsep 2pt
\begin{array}[b]{rclrclrcl}
\alpha(0) &=& 1/137.036, \quad & 
\disp\frac{\pi\alpha(0)}{\sqrt{2}\GF} &=& (37.281\GeV)^2, \quad & 
\sin^2\theta_\PW &=& 0.227, \\
\MZ &=& 93.8\GeV, &  \MH &=& 100\GeV, &
\Mt &=& 180\GeV. 
\end{array}
\eeq
Neither the W-boson mass nor the masses of fermions other than the
top quark are given in \citere{Bardin:1986bc}. 
We supplement the above input parameters by the following fermion masses,
\beq
\arraycolsep 2pt
\begin{array}[b]{rclrclrcl}
M_N & = & m_\Pp = 0.938271998\GeV, \\
\Me &=& 0.5110034\MeV, \quad & m_\mu &=& 0.10565943\GeV, \quad & 
m_\tau &=& 1.7842\GeV,\\
\Mu &=& 80\MeV, & \Mc &=& 1.5\GeV, & \\
\Md &=& 80\MeV, & \Ms &=& 0.30\GeV, & \Mb &=& 4.5\GeV.
\end{array}
\eeq
In detail, the lepton masses are taken from \citere{Roos:sd}
and the quark masses from \citere{Sarantakos:1982bp};
\citere{Bardin:1986bc} indirectly refers to these papers.

The authors of \citere{Bardin:1986bc}
advocate the on-mass-shell renormalization scheme with
the fine-structure constant $\alpha(0)$, the Fermi constant $\GF$,
the Z- and Higgs-boson masses $\MZ$ and $\MH$, and 
all fermion masses as physical input parameters.
In contrast to the conventional on-shell scheme the W-boson mass $\MW$
is not an input parameter but is instead determined
by (iterative) solution of the implicit relation \refeq{eq:deltar}.
Using the above input parameters, except for $\sin^2 \theta_W$,
and solving Eq.~\refeq{eq:deltar} for $\MW$ 
we find for the W-boson mass $\MW = 83.807\GeV$, which leads to
$\sw^2=\sin^2\theta_\PW=0.2017$.
This constitutes roughly a 10\% discrepancy to the value 
$\sin^2\theta_\PW=0.227$ given in \citere{Bardin:1986bc};
it is not clear where the latter is actually used in 
\citere{Bardin:1986bc}.
Having specified $\MW$, we further assume that the authors
of \citere{Bardin:1986bc} proceed as in our $\GF$ scheme, i.e.\
by taking $\GF$ and all particle masses as input.

Concerning the treatment of initial-state quark-mass singularities,
there is a conceptual difference between the calculation of
\citere{Bardin:1986bc} and ours. The authors of \citere{Bardin:1986bc}
do not apply $\overline{\mbox{MS}}$ factorization, as we do, but
instead set the masses $m_q$ of the incoming quarks to
$m_q=x M_N$ without any additional subtractions.
Moreover, the (rather old set of) parton distribution functions used 
in \citere{Bardin:1986bc} are not available to us; the corrections, 
however, should depend on the parton densities only very weakly.

In \refta{tab:BDcomp} we compare our results with the ones quoted
in \citere{Bardin:1986bc}. Since the input is not completely clear,
we use different setups. Specifically, we apply the usual $\GF$-scheme
(denoted ``$\GF$''), where $\MW=82.469\GeV$ is calculated from 
$\sin^2\theta_\PW=0.227$ and $\MZ$,
as well as a modified $\GF$ scheme [denoted ``$\GF$ ($\MW$ derived)'']
that differs from the former in the value of $\MW=83.807$, which is 
calculated from $\GF$ and the particle masses via $\De r$ at one loop. 
\begin{table}
\centerline{
$\begin{array}{cccccc}
\multicolumn{6}{l}{\mbox{Hadronic energy cut:} \quad
E_{\had}^{\LAB} > 10\GeV}
\\ 
\hline 
\mbox{IPS} & \mbox{FS} &
R^\nu_0 & \delta R^\nu_\NC & \delta R^\nu_\CC & \De\sin^2\theta_\PW
\\
\hline 
\multicolumn{1}{c}{\mbox{result of \citere{Bardin:1986bc}}} & \mbox{BD}
	& - & -0.0021  & -0.0223  & -0.0114
\\
$\GF$ & \overline{\mbox{MS}}
	& 0.31455(1) & 0.0010(1)  & -0.0202(4)  & -0.0090(2)
\\
$\GF$\mbox{($\MW$ derived)} & \overline{\mbox{MS}}
	& 0.33113(2) & -0.0018(1)  & -0.0186(4)  & -0.0095(2)
\\
$\GF$ & \mbox{BD}
	& 0.31455(1) & -0.0026(1)  & -0.0184(4)  & -0.0098(2)
\\
$\GF$\mbox{($\MW$ derived)} & \mbox{BD}
	& 0.33113(2) & -0.0050(1)  & -0.0169(4)  & -0.0103(2)
\\
\hline 
\end{array}$ }
\vspace{1em}
\centerline{
$\begin{array}{cccccc}
\multicolumn{6}{l}{\mbox{Hadronic+photonic energy cut:} \quad
E_{\had+\phot}^{\LAB} > 10\GeV}
\\ 
\hline 
\mbox{IPS} & \mbox{FS} &
R^\nu_0 & \delta R^\nu_\NC & \delta R^\nu_\CC & \De\sin^2\theta_\PW
\\
\hline 
$\GF$ & \overline{\mbox{MS}}
	& 0.31455(1) & 0.0018(1)  & -0.0294(4)  & -0.0130(2)
\\
$\GF$\mbox{($\MW$ derived)} & \overline{\mbox{MS}}
	& 0.33113(2) & -0.0010(1)  & -0.0271(4)  & -0.0132(2)
\\
$\GF$ & \mbox{BD}
	& 0.31455(1) & -0.0018(1)  & -0.0276(4)  & -0.0138(2)
\\
$\GF$\mbox{($\MW$ derived)} & \mbox{BD}
	& 0.33113(2) & -0.0043(1)  & -0.0253(4)  & -0.0139(2)
\\
\hline 
\end{array}$ }
\vspace{1em}
\centerline{
$\begin{array}{cccccc}
\multicolumn{6}{l}{\mbox{Hadronic energy cut after $\ga$ recombination:} \quad
E_{\had,\recomb}^{\LAB} > 10\GeV}
\\ 
\hline 
\mbox{IPS} & \mbox{FS} &
R^\nu_0 & \delta R^\nu_\NC & \delta R^\nu_\CC & \De\sin^2\theta_\PW
\\
\hline 
$\GF$ & \overline{\mbox{MS}}
	& 0.31455(1) & 0.00138(2)  & -0.02149(2)  & -0.00943(1)
\\
$\GF$\mbox{($\MW$ derived)} & \overline{\mbox{MS}}
	& 0.33113(2) & -0.00139(2)  & -0.01970(2)  & -0.00988(1)
\\
$\GF$ & \mbox{BD}
	& 0.31455(1) & -0.00214(2)  & -0.01969(2)  & -0.01023(1)
\\
$\GF$\mbox{($\MW$ derived)} & \mbox{BD}
	& 0.33113(2) & -0.00465(2)  & -0.01805(2)  & -0.01064(1)
\\
\hline 
\end{array}$ }
\caption{Comparison with results of \citere{Bardin:1986bc}.}
\label{tab:BDcomp}
\end{table}
Moreover, we apply two different factorization schemes, the
$\overline{\mbox{MS}}$ scheme and the one used in 
\citere{Bardin:1986bc} (quoted as ``BD'').
Of course, this comparison of results can neither prove nor
disprove the correctness of the results of \citere{Bardin:1986bc}.
However, the first part of the table ($E_{\had}^{\LAB} > 10\GeV$)
suggests significant differences in the correction $\De\sin^2\theta_\PW$.
In any case, the variations in the corrections that are due to the
different factorization schemes ($\overline{\mbox{MS}}$ versus BD)
and due to different ways of including the final-state photon
in the hadronic energy in the final state can be as large as
the accuracy in the NuTeV experiment, which is about 0.0016 in
$\sin^2\theta_\PW$ (if statistical and systematic errors are
combined quadratically).

\section{Summary}
\label{se:summary}

\begin{sloppypar}
Results from a new calculation of electroweak ${\cal O}(\alpha)$
corrections to neutral- and charged-current neutrino deep-inelastic
scattering have been presented. The salient features of the calculation
and some delicate points of the corrections, such as the issue of
collinear fermion-mass singularities, have been discussed in detail.
Unfortunately a tuned comparison to older results,
which were used in the NuTeV data analysis, is not possible.
However, a comparison based on an 
assumption for missing input parameters
leaves room for speculating on possible significant differences in the
electroweak radiative corrections.
Therefore, an update of the NuTeV analysis seems to be desirable.
We provide a Fortran code for the electroweak radiative corrections
that could be used in this task.
Specifically, our investigation of the factorization-scheme dependence
for initial-state radiation and of different ways to treat photons in the
final state reveals that these effects can be as large as the
$3\sigma$ difference between the NuTeV measurement and the Standard Model
prediction in the on-shell weak mixing angle.

A careful re-assessment of all theoretical uncertainties involved
in the NuTeV analysis should be worked out, based on the new studies of
both QCD \cite{McFarland:2003jw,Davidson:2001ji}
and electroweak corrections.
The NuTeV collaboration estimated the theoretical uncertainty due to
missing higher-order effects to 0.00005 and 0.00011 in
$\delta R^\nu$ and $\Delta \sin^2 \theta_W$, respectively.
The results on electroweak corrections presented in this paper
indicate that these numbers are too optimistic.
\end{sloppypar}


\section*{Acknowledgments}

One of the authors (K.~D.) would like to thank Michael Spira and 
Ansgar Denner for useful discussions.
Dima Bardin is gratefully acknowledged for providing further information
on \citere{Bardin:1986bc}.
This work was supported in part by the European Union under contract
HPRN-CT-2000-00149.


\end{document}